\documentclass[final,5p,times,twocolumn]{elsarticle}
\journal{New Astronomy}

\usepackage{color}
\usepackage{natbib}

\usepackage{graphicx}	% Including figure files
\usepackage{amsmath}	% Advanced maths commands
\usepackage{amssymb}	% Extra maths symbols
\usepackage{gensymb}

\usepackage{txfonts}
\usepackage{url}
\usepackage{natbib}
\usepackage[utf8]{inputenc}
\usepackage{subfigure}
\usepackage{float}

\def\kms{{\,\rm km\,s^{-1}}}
\def\mpch{\rm \ h^{-1}\rm Mpc}
\def\eone{{\bf e}_{1}}
\def\etwo{{\bf e}_{2}}
\def\ethree{{\bf e}_{3}}
\def\lone{\lambda_{1}}
\def\ltwo{\lambda_{2}}
\def\lthree{\lambda_{3}}
\def\lth{\rm \lambda_{th}}
\def\Rs{\rm R_{\rm s}}

\def\msun{\,\rm M_{\odot}}
\def\hessg{\, \rm \textbf{H}_{\rm ij, g}}
\def\hesshalo{\, \rm \textbf{H}_{\rm Halo}}

\def\loneg{\lambda_{\rm 1, NGP}}
\def\ltwog{\lambda_{\rm 2, NGP}}
\def\lthreeg{\lambda_{\rm 3, NGP}}
\def\lonehalo{\lambda_{\rm 1, Halo}}
\def\ltwohalo{\lambda_{\rm 2, Halo}}
\def\lthreehalo{\lambda_{\rm 3, Halo}}
\def\eoneg{{\bf e}_{\rm 1, NGP}}
\def\etwog{{\bf e}_{\rm 2, NGP}}
\def\ethreeg{{\bf e}_{\rm 3, NGP}}
\def\eonehalo{{\bf e}_{\rm 1, Halo}}
\def\etwohalo{{\bf e}_{\rm 2, Halo}}
\def\ethreehalo{{\bf e}_{\rm 3, Halo}}

\def\eeone{ {\bf e}_{\rm 1, i} \cdot {\bf e}_{\rm 1, j}}
\def\eetwo{ {\bf e}_{\rm 2, i} \cdot {\bf e}_{\rm 2, j}}
\def\eethree{ {\bf e}_{\rm 3, i} \cdot {\bf e}_{\rm 3, j}}
\def\ab{(128,256)}
\def\ac{(128,512)}
\def\bc{(256,512)}

\newcommand{\blue}{\textcolor{blue}}
\newcommand{\red}{\textcolor{red}}
\begin{document}

\begin{frontmatter}

\title{A robust determination of halo environment in the cosmic field}
%\title{More accurate determination of halo environment in the cosmic field}
%\title{An improved algorithm of classification of the cosmic web}
%\tnotetext[mytitlenote]{Fully documented templates are available in the elsarticle package on \href{http://www.ctan.org/tex-archive/macros/latex/contrib/elsarticle}{CTAN}.}

%% Group authors per affiliation:
\author[a,b]{Peng Wang\corref{mycorrespondingauthor}}
%\address{Instituto de Astronom\'ia, Universidad Nacional Aut\'onoma de M\'exico, Ciudad Universitaria, 04510, M\'exico.}
%\fntext[myfootnote]{Since 1880.}
\cortext[mycorrespondingauthor]{Corresponding author}
\ead{wangp410828@gmail.com}

\author[a,c]{Xi Kang}
\ead{kangx@pmo.ac.cn}
\author[b,d]{Noam I. Libeskind}
\author[e]{Quan Guo}
\author[b]{Stefan Gottl\"ober}
\author[a,f]{Wei Wang}

\address[a]{Purple Mountain Observatory, No. 8 Yuan Hua Road, 210034 Nanjing, China}
\address[b]{Leibniz-Institut f\"ur Astrophysik Potsdam,  An der Sternwarte 16, 14482 Potsdam, Germany}
\address[c]{Zhejiang University-Purple Mountain Observatory Joint Research Center for Astronomy, Zhejiang University, Hangzhou 310027, China}
\address[d]{University of Lyon; UCB Lyon 1/CNRS/IN2P3; IPN Lyon, France}
\address[e]{Shanghai Astronomical Observatory, Nandan Road 80, Shanghai 200030, China}
\address[f]{School of Astronomy and Space Science, University of Science and Technology of China, Hefei 230026, Anhui, China}

\begin{abstract}
A number of methods for studying the large-scale cosmic matter distribution exist in the literature. One particularly common method employed to define the cosmic web is to examine the density, velocity or potential field. Such methods are advantageous since a Hessian matrix can be constructed whose eigenvectors (and eigenvalues) indicate the principal directions (and strength) of local collapse or expansion. Technically this is achieved by diagonalizing the Hessian matrix using a fixed finite grid. The resultant large-scale structure quantification is thus inherently limited by the grid's finite resolution. Here, we overcome the obstacle of finite grid resolution by introducing a new method to determine halo environment using an adaptive interpolation which is more robust to resolution than the typical ``Nearest Grid Point''  (NGP) method. Essentially instead of computing and diagonalizing the Hessian matrix once for the entire grid, we suggest doing so once for each halo or galaxy in question.  We examine how the eigenvalues and eigenvector direction's computed using our algorithm and the NGP method converge for different grid resolutions, finding that our new method is convergent faster. Namely changes of resolution have a much smaller effect than in the NGP method. We therefore suggest this method for future use by the community.
\end{abstract}

\begin{keyword}
large-scale structure of Universe; cosmic web; dark matter halo; simulation.
\end{keyword}

\end{frontmatter}

\section{Introduction}\label{sec:intro}
The large-scale structure of the Universe is formed via gravitational instability from the initial seeding of perturbations in the otherwise homogenous density field. On large scales, the matter distribution of the Universe is not uniform, but exhibits a web-like structure, which can be well described by the linear theory and the Zel’dovich approximation\citep{1970A&A.....5...84Z}. The multi-scale web-like structure, commonly referred to as ``the cosmic web''\citep{1996Natur.380..603B}, is well studied and has been been described as a cellular system \citep{1978IAUS...79..241J}. Analyses of large galaxy surveys, such as the 2dF Galaxy Redshift Survey \citep[][]{2003astro.ph..6581C}, the Sloan Digital Sky Survey \citep[][]{2004PhRvD..69j3501T} and the Two Micron All Sky Survey (2MASS) redshift survey \citep{2005ASPC..329..135H} have shown that the cosmic web can be decomposed classified into four categories, namely knots (sometimes referred to as clusters), filaments, walls (sometimes referred to as sheets) and voids. As these terms suggest, in general, knots are formed at the relatively denser regions which sit at the intersection of filaments and are fed by mass flowing along the spines filaments. Filaments are formed at the intersection of walls. Walls are formed abutting voids, which remain relatively under-dense. In general the classification of the cosmic web according to the above hierarchy (namely knots, filaments, sheets and voids) can be done by examining the rate of compression (or expansion) of cosmic material along the three orthonormal axes\citep{1991QJRAS..32...85I, 2014MNRAS.441.2923C, 2018MNRAS.473.1562W}.

One of the original intentions of developing cosmic web classification methods is to understand whether, and if so how, the cosmic web influences the properties of galaxies and the evolution of galaxies within it. Clear evidence has been presented that properties of haloes/galaxies (such as shape, spin and satellite spatial distribution) correlate with their large scale environments \citep{2007ApJ...655L...5A, 2014MNRAS.440L..46A, 2007MNRAS.381...41H, 2007MNRAS.375..489H, 2010MNRAS.405..274H, 2009ApJ...706..747Z, 2015ApJ...798...17Z, 2017MNRAS.468L.123W, 2018MNRAS.473.1562W, 2015MNRAS.446.1458M, 2012MNRAS.421L.137L, 2013ApJ...766L..15L, 2014MNRAS.443.1274L, 2016MNRAS.457..695P, 2015ApJ...800..112G, 2013ApJ...775L..42T, 2015MNRAS.450.2727T, 2017A&A...599A..31H}. In order to understand how the large scale structure and the immediate environment of a halo/galaxy affects its formation and evolution, it is crucial to robustly identify and quantify this large scale environment (LSE) at the position of each halo or galaxy.

In the past two decades, a variety of methods have been devised to classify the cosmic environment based on local variations (density, gravitational potential and velocity) of the matter distribution. \citep{2007ApJ...655L...5A, 2007MNRAS.381...41H, 2014MNRAS.443.1090F, 2012MNRAS.421L.137L, 2014MNRAS.441.2923C}. The reader is also referred  to \cite{2018MNRAS.473.1195L} for the comparison of 12 different methods. Here we focus on the computation of \textit{Hessian based} methods that employ a fixed grid and assign haloes to the Nearest Grid Point (NGP). These methods are frequently used in the analysis of simulations or reconstructions \citep{2015MNRAS.452.1052L} or wherever there is a continuous cosmic field.  \cite{2007MNRAS.381...41H} first proposed an dynamic classification scheme (often referred as T-web or P-web) of the cosmic web based on counting the positive number  of the eigenvalues of the tidal tensor, i.e, the Hessian of gravitational potential. Studies argue that the value of the properly normalized threshold should be around unity \cite{2014MNRAS.443.1090F} or zero \citep{2007MNRAS.381...41H}. Similarly, \cite{2012MNRAS.425.2049H} and \cite{ 2012MNRAS.421L.137L} forwarded the V-web technique based on the signature of the velocity shear field. Instead of using the tidal or velocity shear field configuration, some works also define large scale environment based on the density field itself \citep{2007ApJ...655L...5A, 2014MNRAS.441.2923C, 2009ApJ...706..747Z, 2017MNRAS.468L.123W, 2018MNRAS.473.1562W}.

These \textit{Hessian-NGP-based} methods have been used in numerous studies which are related to the correlation between haloes/galaxies and cosmic web. However there is still much room for improvement. Although some methods discretize cosmic fields using adaptive non-regular grids \citep[i.e. a Delaunay tessellation][]{2014MNRAS.441.2923C}, many still rely on regular grids, such as a ``cloud-in-cell'' algorithm. In those methods which employ a regular grid, the environment properties of a halo/galaxy is often replaced by the environment properties of the nearest grid point \citep[][]{1995pmtn.rept.....M}  without considering the contribution from other neighboring grid points. In other words, these methods discretize a given field and compute the cosmic web at each grid point. A given value/type for the cosmic web can then be assigned to a halo or galaxy which is in the same voxel (``volume pixel'' or grid cell).  Such methods may not be robust since a galaxy has an extended size and mass (i.e. is not simply a point) that will be affected by nearby grid points. In the standard NGP method, environment properties of a halo might  change when the grid size changes, since they are related to the grid and not to the halo. This may be risky if one considers the time evolution of a halo since haloes move relative to the LSE and therefore occasionally they will cross the border of a grid cell. In order to avoid these issues, we introduce an improved algorithm for the computation of Hessian fields by considering the influence from nearby grid points. Our approach consists of three free parameters, the resolution $\rm N_{\rm grid}^{3}$, the smoothing length $\Rs$ and the threshold $\lth$ used to classify the cosmic web. In most works, they mainly focus on $\Rs$ \citep{2007ApJ...655L...5A, 2014MNRAS.441.2923C} and $\lth$ \citep{2014MNRAS.443.1090F}, while less discussed is $\rm N_{\rm grid}^{3}$.

Our paper is organized as follows. Section~\ref{sec:algorithm} presents the algorithm in detail of how our approach works. In Section~\ref{sec:simu}, we introduce the test simulation we used. Our main results are presented in Section~\ref{sec:stab} by testing the stability of the six key parameters of large scale structure of haloes in our approach, and comparing with traditional one. Conclusions and discussion are presented in Section~\ref{sec:con}.

\section{Web classification: algorithm}\label{sec:algorithm}
The algorithm presented in this paper is similar to that suggested by \cite{2007MNRAS.381...41H} but uses the density field instead of the potential. We explain our algorithm in the context of cosmological simulations. We note that our algorithm is not limited to the Hessian of the density field, but could also be applied to any hessian method computed on a regular grid (i.e. such as the potential or the velocity field). I

The algorithm proceeds by computing the smoothed density field $\rho_{s}(\boldsymbol{x})$ at each point in the volume under consideration. First we construct a discrete density field, $\rho_{cic}(\boldsymbol{x})$, from the discrete distribution of particles in the simulation by using the  Cloud-in-Cell (CIC) technique with a given number of grid, $\rm N_{\rm grid}^{3}$. A spherically symmetric Gaussian window function is then applied to smooth this density field to obtain a smoothed, continuous density field $\rho_{s}(\boldsymbol{x})$. The Hessian matrix $\hessg$ is then constructed at each grid point. It is defined as
%   Hessian with density field
\begin{equation}\label{equ:hessian}
\textbf{H}_{ij} = \frac {\partial^2\rho_s(\boldsymbol{x})}
  {\partial x_i \partial x_j},
\end{equation}
where $\rho_s(\boldsymbol{x})$ is the smoothed density field with smoothing length $\Rs$. $i$ and $j$ denote the Hessian matrix indices with values of 1, 2, or 3 corresponding to the three cartesian axes $x,~y,~z$. The Hessian matrix may be diagonalized and its eigenvalues may be sorted  such that ($\lone < \ltwo < \lthree$) and the corresponding eigenvectors  marked as $\eone$, $\etwo$ and $\ethree$, respectively. The number of eigenvalues  bigger than the threshold $\lth$ is used to classify the type of environments of a halo as follows:
\begin{enumerate}
\item knot: if $\lthree < \lth$
\item filament: if $\ltwo < \lth < \lthree$
\item wall: if $\lone < \lth < \ltwo$
\item void: if $\lth < \lone$
\end{enumerate}

In fact, this is the typical way of classifying LSE namely the methodology we wish to improve. Therefore, a new Hessian matrix, $\hesshalo$, is constructed at the location of each halo by considering the contribution of $\textbf{H}_{ij}$ from the nearby 8 grid points weighted according to a CIC scheme. In the process, a halo is regarded as a uniform distribution of a mass cloud . In the following, we give an introduction on each step of the algorithm and give the details.

%  subsection
\subsection{Step 1 - Constructing CIC density field }
\begin{figure} 
\includegraphics[width=0.45\textwidth]{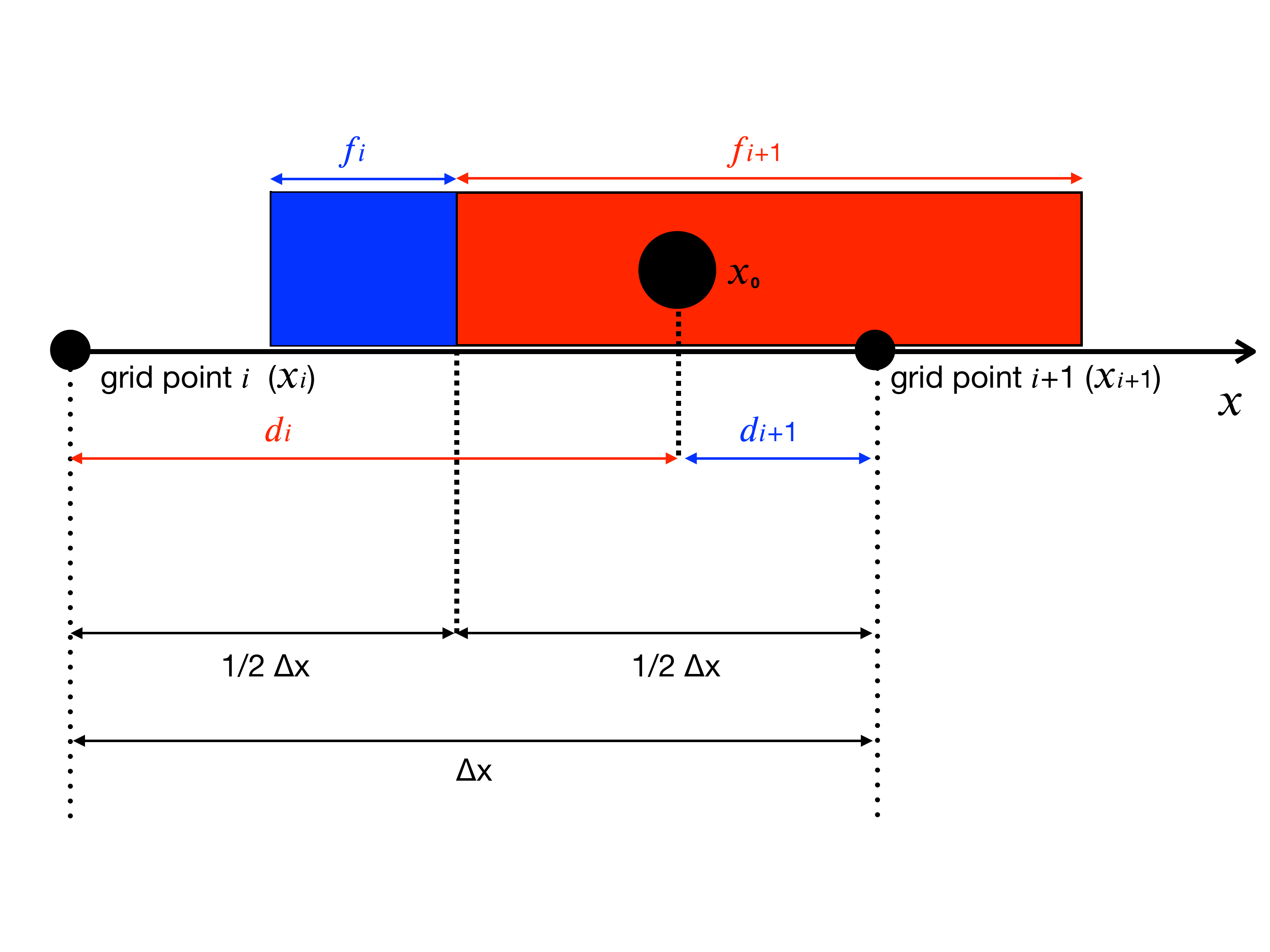}
%\vspace{4cm}
\caption{Mass assignment according to the CIC in 1D. The distance between the test particle and grid point $i$ ($i$+1) is $d_{i}$ ($d_{i+1}$). The different color areas indicate the relative proportions of the mass of a given particle assigned to grid point $i$ (in blue with  $f_{i}$) and $i$+1 (in red with $f_{i+1}$).}
\label{fig:cic}
\end{figure}

For simplicity, here we just use a one-dimensional case to illustrate how to construct the CIC density field from simulation. As shown in the Fig.~\ref{fig:cic}, we show a one-dimensional grid with grid size $\Delta_{x}$. The grid size $\Delta_{x}=L_{box}/N_{grid}=x_{i+1}-x_{i}$, where $L_{box}$ is the simulation box size and $N_{grid}$ is number of grid cells. In a simulation, periodic boundary conditions need to be considered during the CIC process. A test particle with mass $m$ and position $x_{0}$ is located between the $i$-th grid point and the $i$+1-th grid point. In this case the particle is regarded as a uniform distribution of a mass cloud, with width $w=f_{i}+f_{i+1}$, centered about the nominal location $x_{0}$. Usually the mass cloud width $w$ is set equal to the grid size $\Delta_x$. The distance between the test particle and grid point $i$ ($i+1$) is $d_{i}=x_{0}-x_{i}$ ($d_{i+1}=x_{i+1}-x_{0}$). It is easy to prove that $f_{i}$=$d_{i+1}$ and $f_{i+1}$=$d_{i}$. The mass of the particle assigned to grid point $i$ equals to $m_{i}=m \times f_{i}=m \times d_{i+1}$, and to grid point $i$+1 quals to $m_{i+1}=m \times f_{i+1}=m\times d_{i}$. By extending this paradigm to 3D space and repeating it for all particles, we can get the CIC density field $\rho(\boldsymbol{x})$. The CIC density reads
%   equation
\begin{equation}
\rho_{cic}(\boldsymbol{x}) = \sum_{i}^{N} m_{i} \times f(i, \boldsymbol{x})
\end{equation}
here $\boldsymbol{x}$ indicates the location of the grid point containing particle $i$, $m_{i}$ is the mass of $i$-th particle, $N$ is the number of particles in simulation, and $f(i, \boldsymbol{x})$ is the mass fraction of $i$-th particle assigned to grid point $\boldsymbol{x}$. For a given particle, its mass will be assigned to nearby 8 grid points in 3D.

\subsection{Step 2 - Smoothing CIC density field} %$\rho_s(\boldsymbol{x})$}.
The smoothed, continuous density field is defined as:
%   equation
\begin{equation}
\rho_{s}(\boldsymbol{x})=\int d\boldsymbol{y}\rho_{cic}(\boldsymbol{x}) G_{\Rs}(\boldsymbol{y}, \boldsymbol{x})
\end{equation}
where $\boldsymbol{x}$ corresponds to the location of a given grid point and the $\rm G_{\Rs}$ denotes a Gaussian window function with smoothing length, $\Rs$, is given by
%   equation
\begin{equation}
G_{\Rs}(\boldsymbol{y}, \boldsymbol{x})=\frac{1}{(2\pi{\Rs^{2}})^{3/2}}\rm exp(-\frac{|\boldsymbol{y}-\boldsymbol{x}|^{2}}{2\Rs^{2}})
\end{equation}
then the smoothed field $\rho_{s}(\boldsymbol{x})$ can be re-written as
%   equation
\begin{equation}\label{equ:roh_k}
\rho_{s}(\boldsymbol{x})=\int \frac{d^{3}k}{(2\pi)^{3}}e^{-k^{2}\Rs^{2}/2}  \hat{\rho}_{s}(\textbf{k}) e^{i\textbf{k} \cdot \boldsymbol{x}}
\end{equation}
where $\hat{\rho}_{s}(\textbf{k})$ is the Fourier transform of the smoothed density field $\rho_{s}(\boldsymbol{x})$.

%  subsection
\subsection{Step 3 - Computing the Hessian matrix $\hessg$}
From Equations (\ref{equ:hessian}) and (\ref{equ:roh_k}), we find the Hessian matrix at each grid point, $\hessg$, reads
%   equation
\begin{equation}
\textbf{H}_{ij, g} = \mathcal{F}^{-1}\{\hat{\textbf{H}}_{ij, g}(\textbf{k})\}
\end{equation}
where $\hat{\textbf{H}}_{ij, g}(\textbf{k})$ denotes the Fourier transform of the Hessian, given by
\begin{equation}
\hat{\textbf{H}}_{ij, g}(\textbf{k}) = -k_{i}k_{j} e^{-k^{2}\Rs^{2}/2}  \hat{\rho}_{s}(\textbf{k})
\end{equation}

%  subsection
\subsection{Step 4 - Converting $\hessg$ to $\hesshalo$}
The advantage of our algorithm is that we convert the Hessian matrix on the grid point, $\hessg$,  to the Hessian matrix $\hesshalo$, corresponding to the position of each halo. We fully consider the influence of the surrounding 8 grid points to a given halo. The $\hesshalo$ is given by
\begin{equation}
\hesshalo = \sum_{\alpha, \ \beta, \ \gamma=0,1} \textit{f}_{\alpha\beta\gamma} \times H_{ij, g}^{\alpha\beta\gamma}
\end{equation}
in which $f_{\alpha\beta\gamma}$ is the contributed fraction of nearby 8 grid points to the halo position. The $f_{\alpha\beta\gamma}$ is calculated similar to what shown in Fig.~\ref{fig:cic}, but the test particle is replaced by a given halo. In this case, halo is regarded as a uniform distribution of a mass cloud with width equals to the grid size.

%  subsection
\subsection{Step 5 - Computing eigenvalues and eigenvectors.}
Solving the Hessian matrix $\hesshalo$,  we obtain the eigenvalues ($\lonehalo \leq \ltwohalo \leq \lthreehalo$) and corresponding eigenvectors ($\eonehalo$, $\etwohalo$, $\ethreehalo$) for every halo. In the typical NGP method, the eigenvalues and eigenvectors of any halo is given by those of the NGP. Namely,  haloes with the same NGP inherit the same large scale structure information (eigenvalues and eigenvectors). This is not robust because the NGP large-scale information depends on the resolution  of the grid (namely the cell size). In order to compare our algorithm with typical NGP, we also computing the eigenvalues and eigenvectors of $\hessg$ and assign these to all haloes which are saved as $\loneg$, $\ltwog$, $\lthreeg$ and $\eoneg$, $\etwog$, $\ethreeg$.

\section{Test simulation}\label{sec:simu}
The simulation data used in this work is Illustris-1\citep{2014MNRAS.444.1518V} with full physics and high mass resolution. The Illustris simulation suite consists of a set of cosmological hydrodynamical simulations carried out with the moving mesh code AREPO \citep{2010MNRAS.401..791S} with following cosmological parameters: $\Omega_m=0.2726$, $\Omega_\Lambda=0.7274$, $\Omega_b=0.0456$, $\sigma_8=0.809$, $n_s=0.963$ and $H_0=100 \ h \kms$ with $h=0.704$. These parameters are consistent with the latest Wilkinson Microwave Anisotropy Probe (WAMP)-9 measurements \citep{ 2013ApJS..208...19H}. Illustris-1 covers a cubic cosmological box of $\rm 75 \mpch$ wide, with periodic boundaries, and within which $1820^3$ dark matter particles and $1820^3$ initial gas cells are evolved from $z=127$ to $z=0$.  The mass resolution is $6.26\times10^6 \msun$ in dark matter and $1.26\times10^6 \msun$ in baryonic matter.

Dark matter halos are identified using the standard Friend-of-Friend (FoF) algorithm \citep{1985ApJ...292..371D} with the linking length equals to  0.2 times the mean particle separation. Baryonic particles (stellar particles, gas cells, SMBH particles) were attached to these FOF primaries in a secondary linking stage \citep{2009MNRAS.399..497D}. The minimum particle number per FOF group is 32. With this procedure, 7,713,601 FOF groups  are identified. The halo position is defined as the position of the most bound particle of the biggest gravitationally bound structures identified by the SUBFIND algorithm \citep{2001MNRAS.328..726S, 1985ApJ...292..371D}. Halo virial mass is marked as $\rm M_{vir}$, defined as the mass which is contained in a spherical region with average density 200 times the critical density of the universe.

\section{Results: stability}\label{sec:stab}
In identifying the large-scale environment for each halo, our algorithm requires three free parameters: the resolution $\rm N_{\rm grid}^{3}$, the smoothing length $\Rs$, and the threshold $\lth$. As described in the introduction, a lot of work \citep{2007ApJ...655L...5A, 2014MNRAS.441.2923C, 2014MNRAS.443.1090F} has focused on the effects of the smoothing length and threshold. In this work, we mainly discuss the dependence on the resolution from small to large and check the stability by comparing the results between our algorithm and the NGP method. Therefore, in the following analysis, we fixed the smoothing length and threshold by setting $\Rs\sim2\mpch$ and $\lth=0.0$, which are values typically used previously published articles \citep[i.e.][]{2007MNRAS.381...41H,2007MNRAS.375..489H,2015MNRAS.452.1052L,2015ApJ...813....6K, 2017MNRAS.468L.123W, 2018MNRAS.473.1562W}.

\begin{figure*} 
\includegraphics[width=0.98\textwidth]{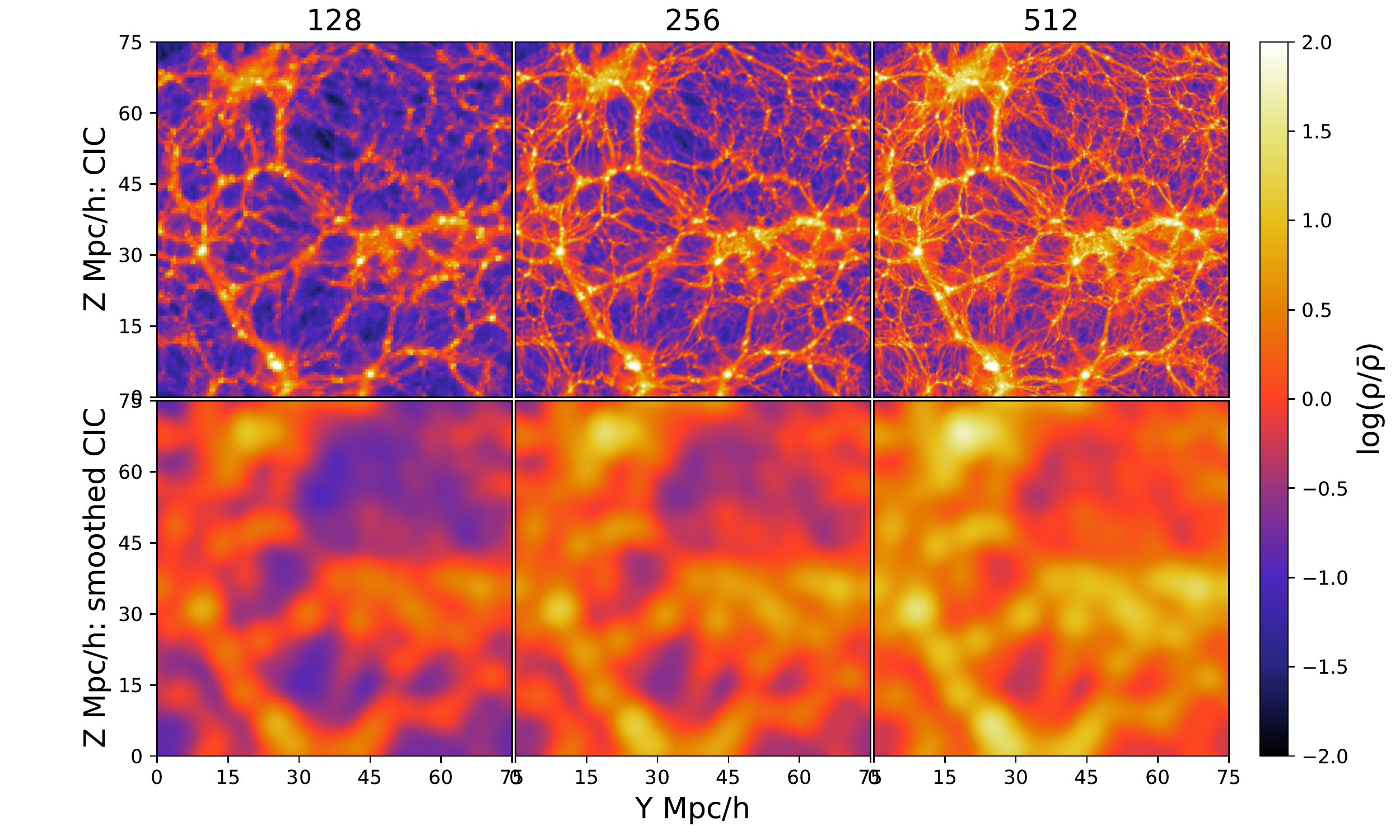}
%\vspace{4cm}
\caption{Upper panels show the logarithm of the normalized CIC density field, $\rm log (\rho/\bar{\rho})=log(1+\delta)$, of a slice of width  $\sim 0.6 \mpch$ across x-axis  (in which $\bar{\rho}$ is the mean density of the universe). Similar to the upper panels, bottom panels show the logarithm of the  CIC density field  when smoothed on a scale $\rm R_s =2 \mpch$, ( see  \textit{Step 2} in Section 2.2 for more detail).  Different columns represent different grid resolutions: $128^3$ in the left,  $256^3$ in the middle and $512^3$ in the right. Values are coded in color bar.}
\label{fig:matter_field}
\end{figure*}
%The logarithm of normalized density field, $\rm log (\rho/\bar{\rho})$, of a slice of width  $\sim 0.6 \mpch$ across x-axis (in which $\bar{\rho}$ is the mean density of whole simulation box). 

\begin{figure*}[!h]
\centering
\subfigure{\includegraphics[width=0.32\textwidth]{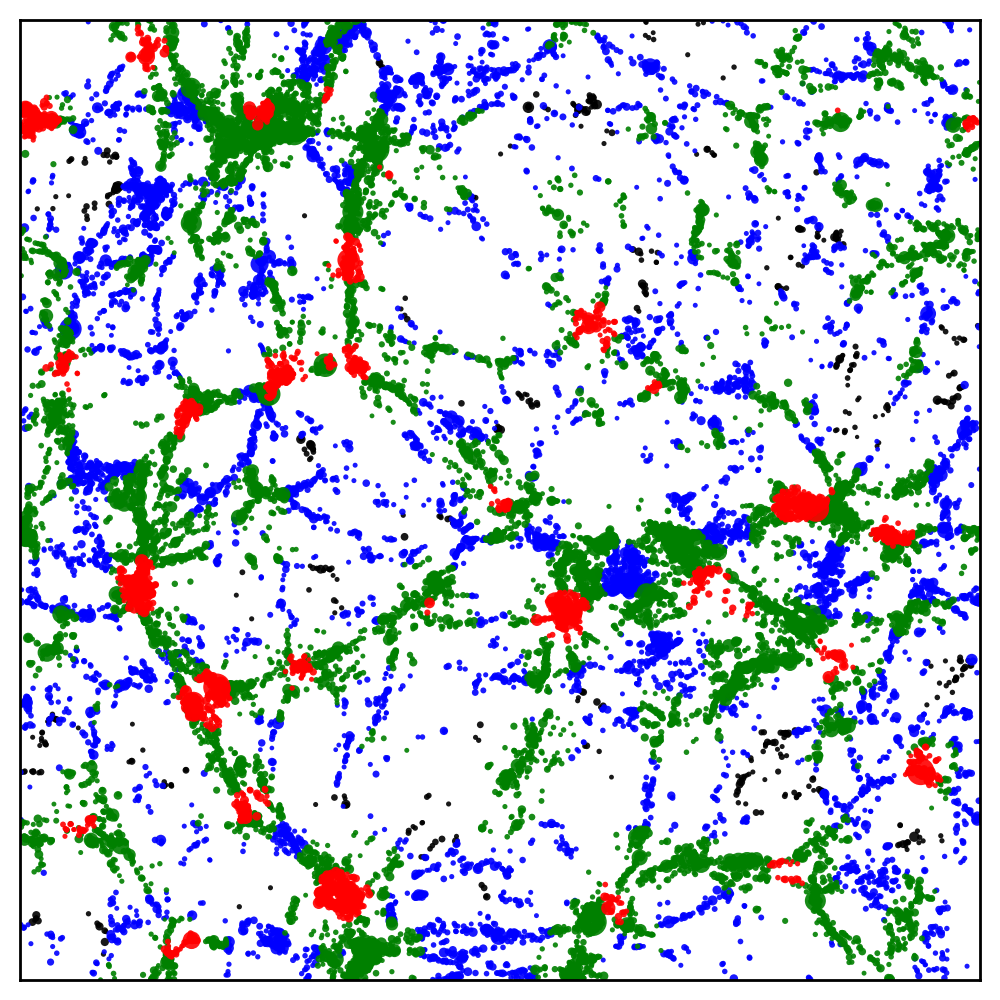}}
\subfigure{\includegraphics[width=0.32\textwidth]{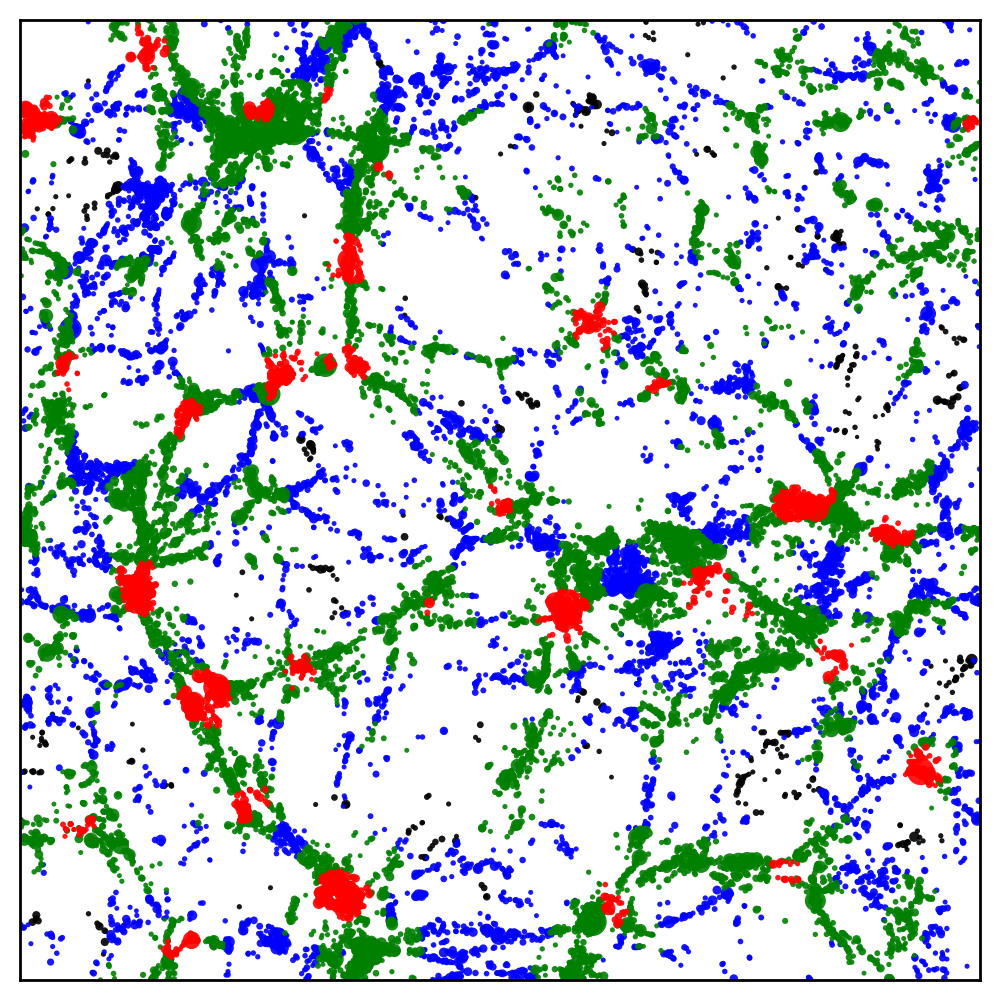}}
\subfigure{\includegraphics[width=0.32\textwidth]{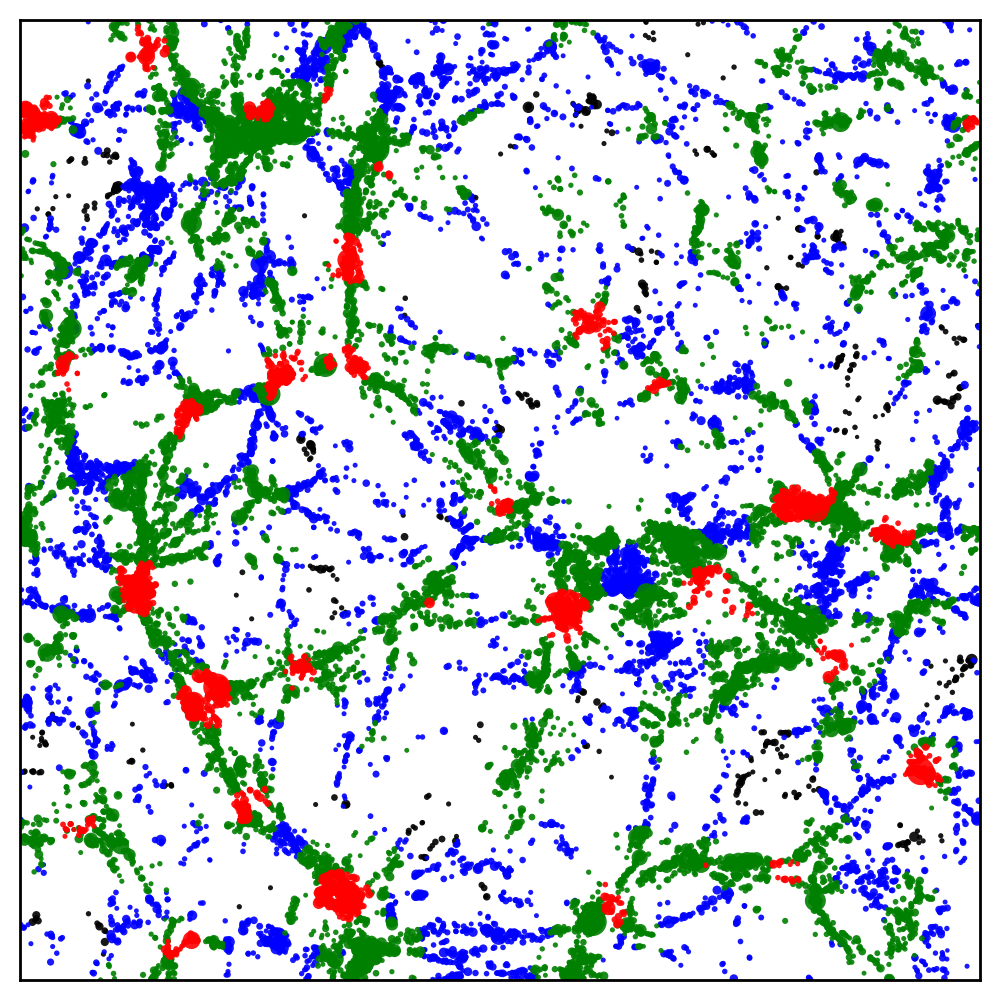}}
\caption{Halo distribution colored by large scale environments classification (in the same slice as Fig.~\ref{fig:matter_field}) according to our algorithm with $128^3$ (left panel), $256^3$ (middle panel) and $512^3$ (right panel) grids. The halos in the four different environments are found in  knots (red), filaments (green), walls (blue), and voids (black) defined by counting the number of positive eigenvectors. These three look almost the same visually, and statistically $\sim98\%$ haloes have the same large scale environment independent of grid size.}
\label{fig:cosmic web}
\end{figure*}

In Fig.~\ref{fig:matter_field} we present a $\sim0.6\mpch$ thick slice the CIC of the density field before (upper panels) and after (bottom panels) applying the Gaussian smoothing. We consider all dark matter particles, gas cells and stars at $z=0$ as the input of \textit{Step 1}. We applied and compared three grid cell numbers $\rm N_{\rm grid}^{3} =128^{3}$, $256^{3}$ and $512^{3}$ and show these in the left, middle and right column, respectively. The corresponding grid resolutions are $\sim0.6\mpch$, $\sim0.3\mpch$ and $\sim0.15\mpch$.   The density field is normalized by the mean, and shown in log $\delta$ (as shown shown in the  color bar). There is a lack of systematic consensus on which smoothing length is the best to characterize the mass distribution on large scales, as this is essentially an arbitrary decision. In most studies, a constant smoothing length ($\sim0.5$ to $\sim2\mpch$) is used at $z=0$ \citep{2007MNRAS.381...41H, 2007MNRAS.375..489H, 2009ApJ...706..747Z, 2013ApJ...762...72T, 2012MNRAS.427.3320C}. \cite{2014MNRAS.441.1974L} argue for a smoothing length that must vary with redshift and is adjusted according to the rms of the density field. For partial discussions, we refer readers to \cite{2007MNRAS.375..489H} and \cite{2014MNRAS.443.1090F} for more details.

Here, we apply the smoothing scale $\Rs$ equals to $2\mpch$. Note that even for other values of $\Rs$ the results do not change much compared to typical NGP method. For both CIC density fields (upper panels) and smoothed density fields (bottom panels), we find that, visually, the skeleton (namely the location of high density regions) looks very similar. However, some difference need to be pointed out. As $\rm N_{\rm grid}^{3}$ increases, more and more fine structure appears in the CIC density filed, shown in upper panels. The under-dense regions shown in the left-upper panel are replaced by some finer structures  whose density may be slightly higher than the mean density in the middle- and right-upper panel. A similar effect is seen for the smoothed density field.

%   figure\aa
\begin{figure*}[!hbpt]
\centerline{\includegraphics[width=0.98\textwidth]{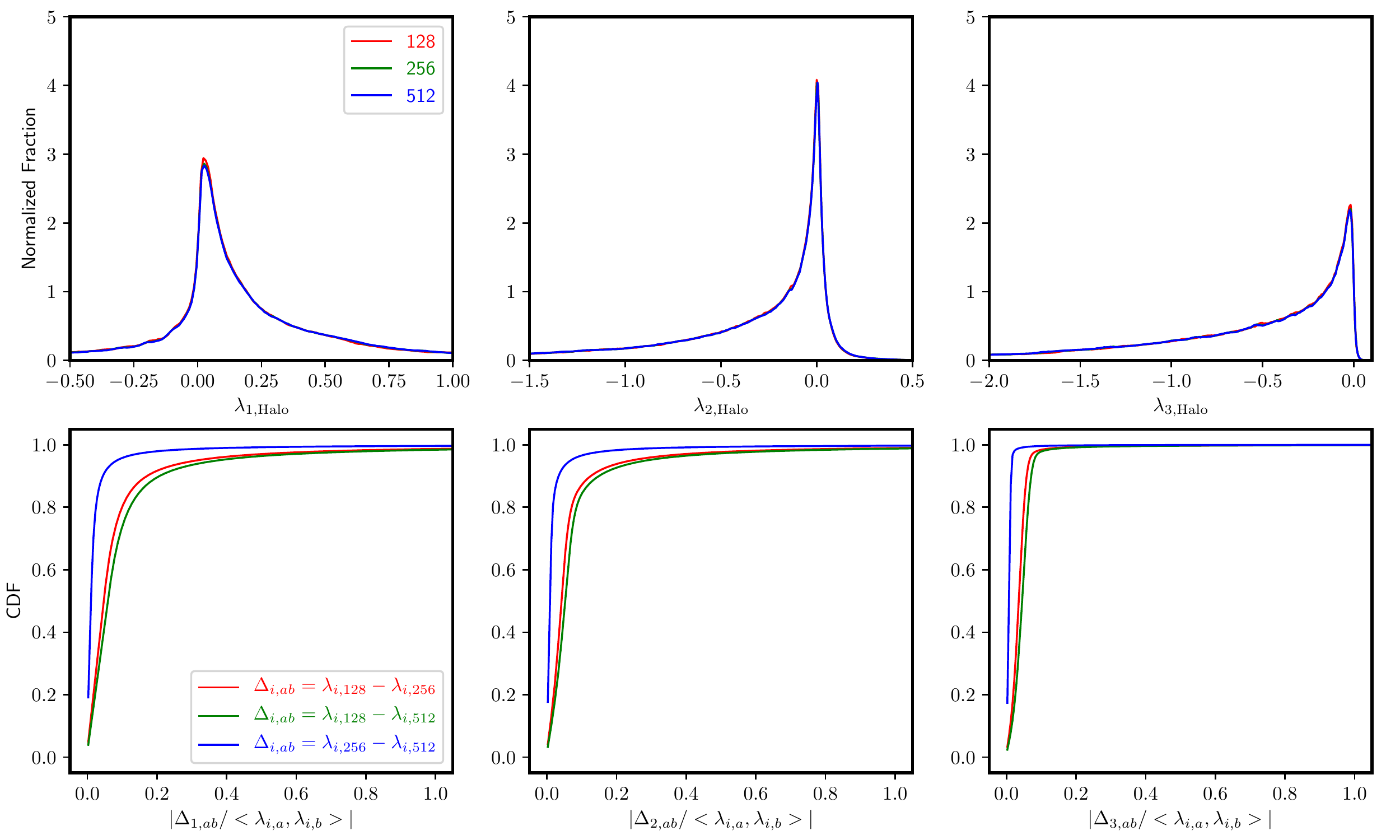}}
\caption{The stability of three eigenvalues ($\lonehalo$, $\ltwohalo$, and $\lthreehalo$) for haloes in our approach with three grid resolutions $\rm N_{\rm grid}^{3}$. Upper panels: a histogram of eigenvalues for $\rm N_{\rm grid}^{3}$, $128^3$ (in red), $256^3$ (in green) and $512^3$ (in blue) grids. Note these are indistinguishable. Bottom panels: the cumulative distribution of the fractional difference in the assignment of eigenvalues, namely  $(\lambda_{i,a}-\lambda_{i,b})/<\lambda_{i,a},\lambda_{i,b}>$, where $a,b$=128, 256, 512 and $i=1,2,3$. Red line indicate the fractional change between eigenvalues calculated by setting $\rm N_{\rm grid}^{3}=128^{3}$ and $\rm N_{\rm grid}^{3}=256^{3}$, green for $128^{3}$ and $512^{3}$, and blue for $256^{3}$ and $512^{3}$.}
\label{fig:lambda_halo}
\end{figure*}
%   figure
\begin{figure*}[!hbpt]
\centerline{\includegraphics[width=0.98\textwidth]{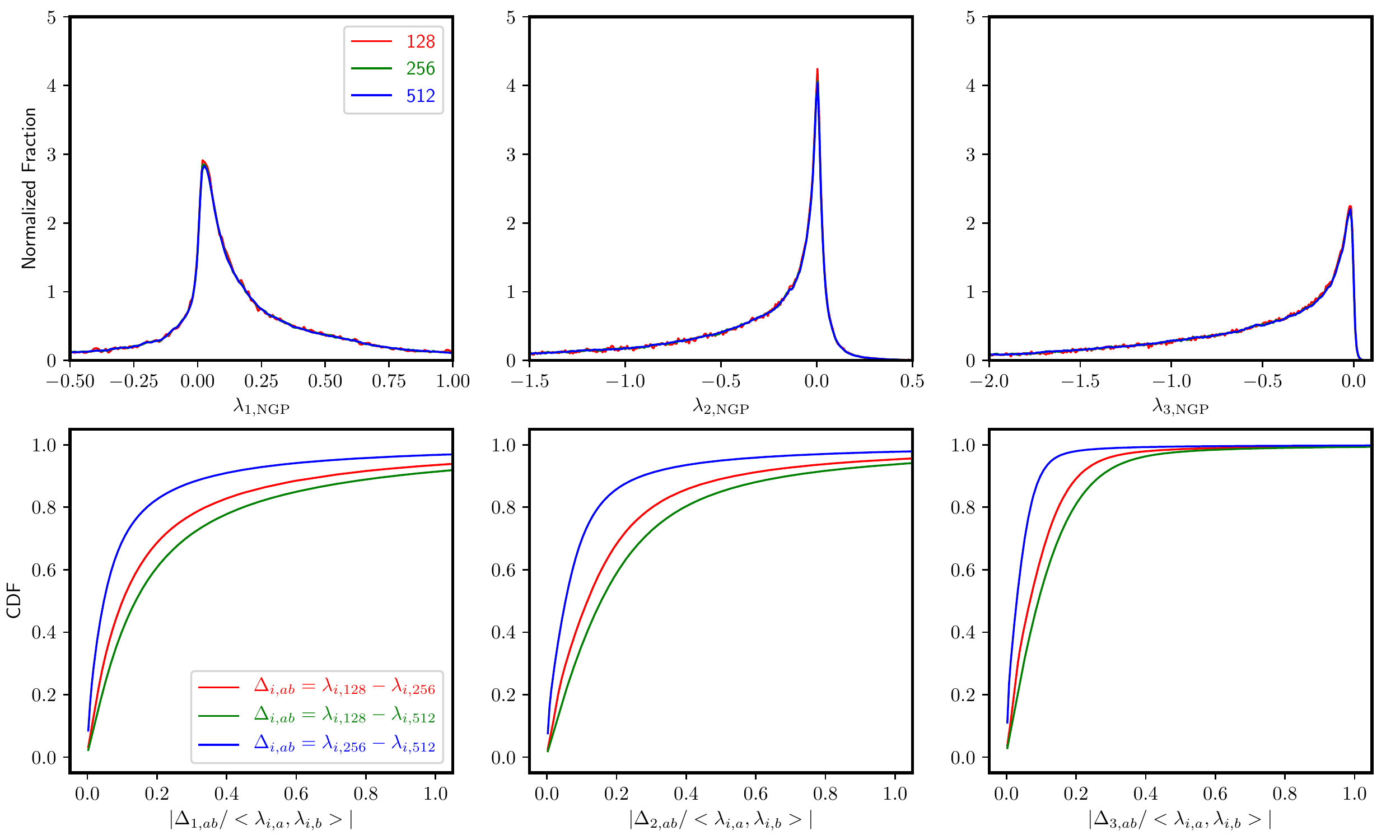}}
\caption{Same as Fig.~\ref{fig:lambda_halo}, but for haloes with eigenvalues ($\loneg$, $\ltwog$ and $\lthreeg$) assigned by NGP. Note that the bins same as Fig.~\ref{fig:lambda_halo}.}
\label{fig:lambda_cell}
\end{figure*}
%   figure
\begin{figure*}[!hbpt]
\centerline{\includegraphics[width=0.98\textwidth]{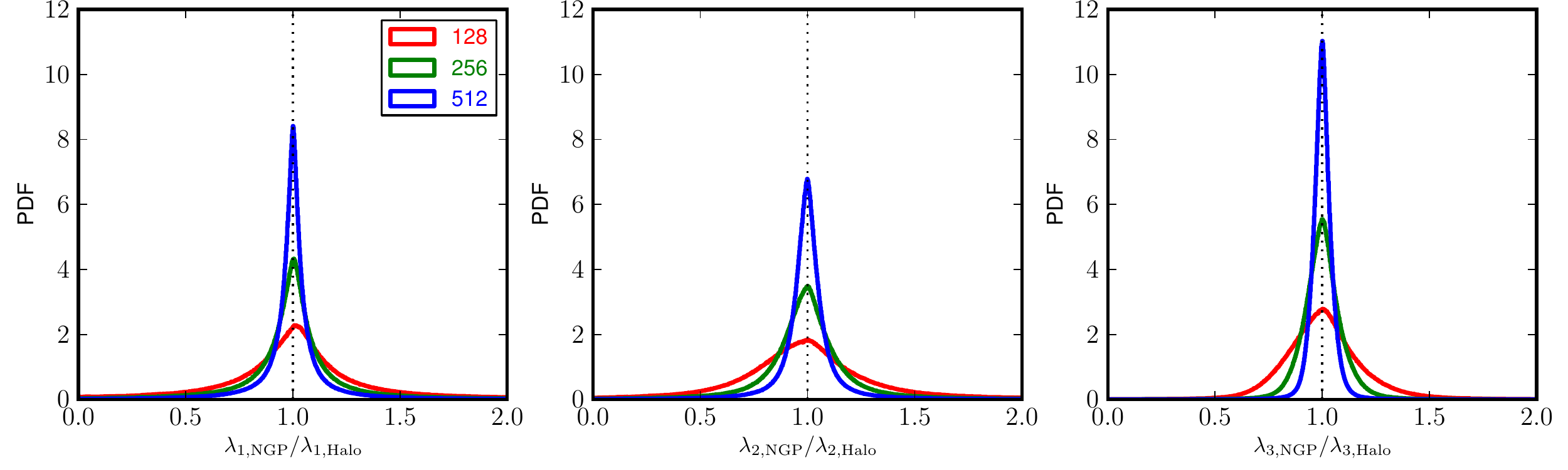}}
\caption{The ratio of three eigenvalues ($\lone$, $\ltwo$ and $\lthree$) assigned to all haloes between our approach (with subscript ``halo'') and the traditional one (with subscript ``NGP''), for three different grid sizes, $128^{3}$ (in red), $256^{3}$ (in green) and $512^{3}$ (in blue).}
\label{fig:lambda_two}
\end{figure*}
%   figure
\begin{figure*}[!hbpt]
\centerline{\includegraphics[width=0.98\textwidth]{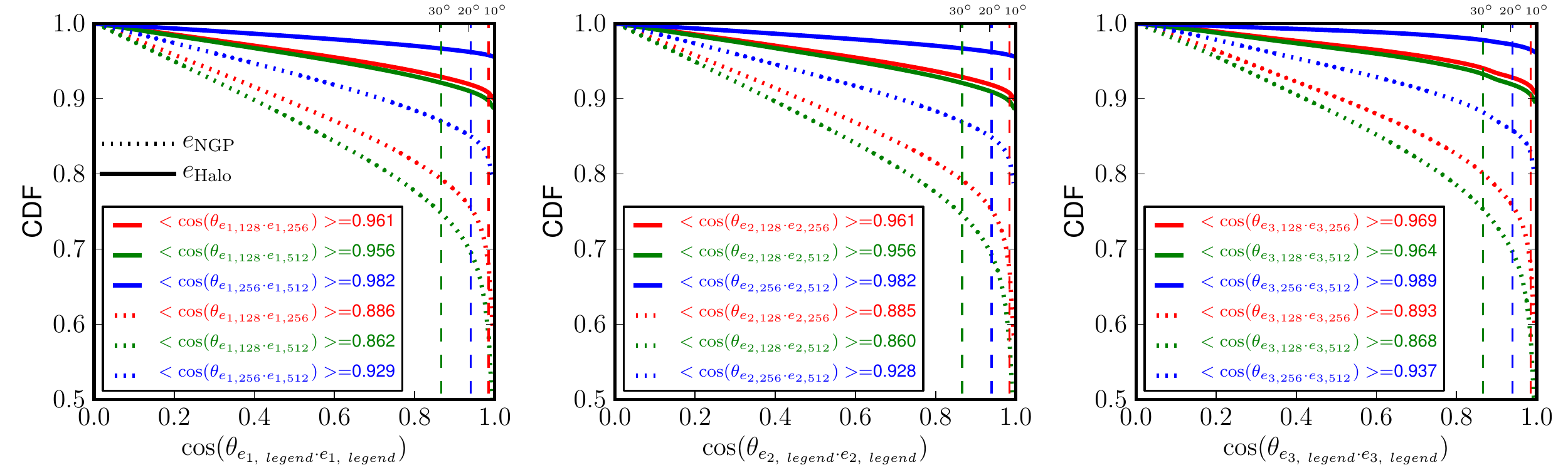}}
\caption{The cumulative distribution function (CDF) of the alignment angle of eigenvectors with changing of grid size $\rm N_{\rm grid}^{3}$ (see legend). Solid lines represent the eigenvectors calculated in our approach ($e_{halo}$), while dotted  lines denote the traditional NGP method ($e_{g}$). Three vertical dashed lines represent  $10^{\circ}$, $20^{\circ}$, and $30^{\circ}$ (in red, blue and green) respectively. The mean $\cos(\theta)$ of each angle are shown in the legend. The  fraction of angles angle smaller than $10^{\circ}$, $20^{\circ}$ and $30^{\circ}$ are shown in Table~\ref{table:cdf_halo} and Table~\ref{table:cdf_ngp}.}
\label{fig:ve_two}
\end{figure*}
%   table
\begin{table*}[!hbpt]
\caption{The fraction of eigenvector alignments smaller than $10^{\circ}$, $20^{\circ}$ and $30^{\circ}$, with our approach, corresponding to solid color lines in Fig.~\ref{fig:ve_two}. The highlighted number in blue show fractions of eigenvectors that are well aligned within $10^{\circ}$ between lowest and highest resolution. Number in red shows the fractions of eigenvectors that are within ($<30^{{\circ}}$) between two grid resolutions.}
\begin{center}
\setlength{\tabcolsep}{0.35mm}{
\begin{tabular}{c|c|c|c|c|c|c|c|c|c} \hline
                          & \multicolumn{3}{c|}{$\eeone$} & \multicolumn{3}{c|}{$\eetwo$} & \multicolumn{3}{c}{$\eethree$} \\ \hline
    $(i, j)=$          &  $\ab$ & $\ac$ & $\bc$ & $\ab$ & $\ac$ & $\bc$ & $\ab$ & $\ac$ & $\bc$  \\ \hline
   $\le10^{\circ}$ & 90.5\% & \blue{89.4\%} & 95.7\% & 90.5\% & \blue{89.3\%} & 95.7\% & 91.4\% & \blue{90.2\%} & 96.4\% \\ \hline
   $\le20^{\circ}$ & 91.8\% & 90.9\% & 96.2\% & 91.7\% & 90.8\% & 96.2\% & 92.7\% & 91.8\% & 97.1\% \\ \hline
   $\le30^{\circ}$ & 92.8\% & 92.1\% & \red{96.6\%} & 92.8\% & 92.0\% & \red{96.7\%} & 94.0\% & 93.2\% & \red{97.8\%} \\ \hline
\end{tabular}
}
\end{center}
\label{table:cdf_halo}
\end{table*}
%   table
\begin{table*}[!hbpt]
\caption{Same as Table~\ref{table:cdf_halo}, but for $\rm e_{\rm NGP}$, corresponding to the dotted color line in Fig.~\ref{fig:ve_two}.}
\begin{center}
\setlength{\tabcolsep}{0.35mm}{
\begin{tabular}{c|c|c|c|c|c|c|c|c|c} \hline
                          & \multicolumn{3}{c|}{$\eeone$} & \multicolumn{3}{c|}{$\eetwo$} & \multicolumn{3}{c}{$\eethree$} \\ \hline
    $(i, j)=$          &  $\ab$ & $\ac$ & $\bc$ & $\ab$ & $\ac$ & $\bc$ & $\ab$ & $\ac$ & $\bc$  \\ \hline
   $\le10^{\circ}$ & 64.8\% & \blue{54.5\%} & 81.1\% & 64.0\% & \blue{53.6\%} & 80.8\% & 65.2\% & \blue{54.8\%} & 81.9\% \\ \hline
   $\le20^{\circ}$ & 75.1\% & 69.0\% & 84.8\% & 74.8\% & 68.7\% & 84.7\% & 75.4\% & 69.1\% & 85.6\% \\ \hline
   $\le30^{\circ}$ & 79.1\% & 74.5\% & \red{86.9\%} & 79.0\% & 74.3\% & \red{86.8\%} & 79.8\% & 75.0\% & \red{88.1\%} \\ \hline
\end{tabular}
}
\end{center}
\label{table:cdf_ngp}
\end{table*}

Its is perhaps obvious that with different grid resolutions and smoothing scales, the appearance of the density field - specifically the characteristics of the density contrast - will vary, since grid resolution and smoothing sets the characteristic scale. This naturally leads to the question of how quantification of the cosmic web changes with scale or grid resolution. In studying the effects of the cosmic web on the properties of dark matter haloes, the identification algorithm should consistently identify the large-scale environment of each halo in some ``converged'' fashion. In other words, the large-scale environment of a given halo is fixed at a given scale. This is natural and expected if the universe is non-fractal, which we believe it is (see \cite{2005ApJ...624...54H}). The question which remains is which scale is relevant for the effects of galaxy formation \color{black}. Furthermore it is incumbent that the large-scale environment of a halo (at fixed scale) should not change with the free parameters of the algorithm, specifically the grid resolution. In the following, we  discuss the stability of the cosmic web classification as function of resolution between our algorithm and the NGP method. The type of the environments, three eigenvalues and eigenvectors of the cosmic web are examined.

By setting the eigenvalue threshold $\lth=0.0$, we may assign each halo one of the four different types of environments. As shown in Fig.~\ref{fig:cosmic web}, we show the halo distribution and their large scale environment classifications of the same slice as Fig.~\ref{fig:matter_field} but with varyong grid sizes. Here halos in the four different cosmic web environments are shown in different colors: knots in red, filaments in green, walls in blue, and voids in yellow. It should be noted that the distribution shown here represents a projected slice of $\sim0.6\mpch$ thickness, along the z-axis.  In this case,  walls appear as roughly as 1D filamentary or isolated structures, while some filaments appear as cluster like.  Despite the visual difference shown in the Fig.~\ref{fig:matter_field} we find that, statistically, $\sim98\%$ of all halos are assigned the same cosmic web environments with our new algorithm; this is compared to 90\% with the NGP method. This doesn't change as function of $\lth$. We examined 10  different none-zero thresholds of $\lth$ from $0.1$ to $1.0$ \citep{2014MNRAS.443.1090F} and the conclusion remains the same: our new algorithm consistently places halos in the same large scale environment irrespective of the underlying grid size.

A deeper comparison is set up to check whether the six parameters (three eigenvalues and three eigenvectors) of the large scale structure assigned to each halo is kept stable at fixed smoothing but with varying grid resolution. The magnitude of an eigenvalues is an important criterion for classifying large-scale environments. Therefore, the stability of eigenvalues in the case of changing parameters has a decisive effect on the classification of large-scale environments. In Fig.~\ref{fig:lambda_halo} and Fig.~\ref{fig:lambda_cell}, we examine the stability of the distribution of eigenvalues both in our algorithm and NGP. The upper panels both in the Fig.~\ref{fig:lambda_halo} and Fig.~\ref{fig:lambda_cell} show the distribution of the three eigenvalues, for $\rm N_{\rm grid}^{3}$ equal to $128^3$ in red, $256^3$ in green and $512^3$ in blue. It can be seen that the distributions look similar. However when we examine the fractional difference with respect to the mean eigenvalue (namely $(\lambda_{i,a}-\lambda_{i,b})/<\lambda_{i,a},\lambda_{i,b}>$, where $a,b$=128, 256, 512 ands $i=1,2,3$), a measure of how significant any difference of the eigenvalue is when changing grid resolution, we see the superiority of the proposed method. This is shown in the bottom panels both in the Fig.~\ref{fig:lambda_halo} and Fig.~\ref{fig:lambda_cell}. Red lines indicate the fractional difference with $\rm N_{\rm grid}^{3}=128^3$ and  $\rm N_{\rm grid}^{3}=256^3$, green lines for $\rm N_{\rm grid}^{3}=128^3$ and $\rm N_{\rm grid}^{3}=512^3$, and blue lines for $\rm N_{\rm grid}^{3}=256^3$ and $\rm N_{\rm grid}^{3}=512^3$. When comparing the 512 resolution to the 128 resolution we find that in our method, $\sim80\%$ of the haloes have eigenvalues within $10\%$ of each other (Fig.~\ref{fig:lambda_halo}), while in the NGP method this is only $58\%$ (Fig.~\ref{fig:lambda_cell}). This indicates that eigenvalues calculated by our approach are relatively much more stable than in the typical NGP method. These numbers are (roughly) independent of which eigenvalue we examine. The smallest eigenvalue is the most stable in both cases. 

In Fig.~\ref{fig:lambda_two}, we show the ratio of the three eigenvalues ($\lone$ in the left panel, $\ltwo$ in the middle panel and $\lthree$ in the right panel) assigned to each halo in our algorithm (with subscript of `halo' ) and the NGP (with subscript of `NGP' ) approach. The vertical dotted line indicates $\lambda_{\rm NGP}=\lambda_{\rm Halo}$, so that the narrower the distribution and the higher the peak at 1.0, indicates that $\lambda_{\rm NGOP}\approx \lambda_{halo}$. It can be seen that as the grid resolution increases, the NGP method approaches our new method. For coarser grids, the agreement is worse.

FInally we further check the stability of the direction of three eigenvectors of cosmic web by checking the alignment of eigenvectors computed at a given scale but with different grid reoslutions. In Fig.~\ref{fig:ve_two}, we show the cumulative distribution function of the alignment angle between eigenvectors ($\eone$ in left panel, $\etwo$ in middle panel and $\ethree$ in the right panel)  calculated with different grid sizes. The alignment angle between the eigenvectors calculated with $\rm N_{\rm grid}^{3}=128^3$ and with $\rm N_{\rm grid}^{3}=256^3$ are shown as red lines, green lines for $\rm N_{\rm grid}^{3}=128^3$ and $\rm N_{\rm grid}^{3}=512^3$, and blue lines for $\rm N_{\rm grid}^{3}=256^3$ and $\rm N_{\rm grid}^{3}=512^3$. Solid  lines represent our algorithm and dotted  lines for NGP. The mean value of the cosine of the alignment angle are shown in the legend in each panel. Three vertical dashed line represents a $10^{\circ}$, $20^{\circ}$ and $30^{\circ}$ angle in red, blue and green respectively.

The solid lines in Fig.~\ref{fig:ve_two}, clearly demonstrate that our new algorithm produces eigenvectors that are well aligned with themselves irrespective of grid resolution, in stark contrast to the NGP case (dotted lines). This indicates that, with different grid resolutions, the eigenvectors calculated by our approach have a higher chance to point in the same direction. This is also seen by examining the mean value of the $\cos(\theta)$. For a quantitative comparison, we show the fraction of eigenvectors aligned within $10^{\circ}$, $20^{\circ}$ and $30^{\circ}$ in the Table~\ref{table:cdf_halo} (our algorithm) and the Table~\ref{table:cdf_ngp} (NGP method). The highlighted number in blue shows the fraction of eigenvectors that are aligned within $10^{\circ}$ between the $\rm N_{\rm grid}=128^{3}$ and $\rm N_{\rm grid}=512^{3}$ case. We find that with our algorithm the fraction reaches up to $90.2\%$ for $e_{3}$ and nearly $90\%$ of all eigenvectors. While shown in Table~\ref{table:cdf_ngp}, this fraction only about $\sim54\%$ for the NGP. Furthermore, the fraction of alignment angle smaller than $30^{\circ}$ of $(256,512)$ (red in Table~\ref{table:cdf_ngp}) is still lower than $(128,512)$ in our algorithm (blue in Table~\ref{table:cdf_halo}).  This indicate that our algorithm can produce eigenvectors whose directions are reliable using different value of $\rm N_{\rm grid}^{3}$.

%%%%%%%%
%  con & dis
%%%%%%%%
%--------------------------------------------------------------------
\section{Conclusion and Discussion}\label{sec:con}
Many studies have investigated the effect of environment on galaxy and halo formation. In cosmological simulations one of the common ways to define environment is via deformation of the matter field. This may be done by examining, for example, the tidal-shear field or the potential field. Such approaches rely on discretising the field, normally using a regular grid. However as particle number in cosmological simulations continues to grow, the  computational resources required to analyse the cosmic web increases as well. The finest grid size that can be used depends on the simulation's resolution  (particle number). This grid must be smoothed in order to dull the discretizaion. The smoothing scale should be the only scale that matters as it sis a physical and not an algorithmic scale. Therefore, In order to obtain robust results, we  require a cosmic web algorithm that is independent of grid size and depends only on the scale of the smoothed field.

We have thus come up with a superior algorithm for computing hessian based cosmic web cosmic web classification (see Section~\ref{sec:algorithm} for details). Most approaches compute the value of a tensor (say the hessian of the density, potential or  tidal field etc) on a regular grid. These are then diagonalized: the eigenvectors and eigenvalues play a key role in the cosmic web quantification. In most such algorithms, halos and galaxies inherit the values of the eigenvectors and eigenvalues from the nearest grid point (NGP) of the grid on which the cosmic web classification was done. Since grid cells can be large, this can lead to inaccuracies. Our improved algorithm essentially computes the hessian at the location of each halo as a weighted mean of the Hessians in the 8 neighboring grids, in a CIC-inspired scheme.

We used the state-of-the-art hydrodynamic simulation, Illustris-1 (z=0), as test data , applying our an improved identification algorithm in order to obtain the cosmic web properties of each halo/galaxy in the simulation.  Comparing with the typical NGP method, we found that our improved algorithm can reach high accuracy even at low resolution. It should be pointed out that our algorithm is not only applicable to the Hessian-based method with the density field, but also applicable to the T-web (tidal fields) and the V-web (velocity field). It can also be applied to those algorithms with other density estimations. Our method produces eigenvectors that don't change direction, and eigenvalues that don't change magnitude with grid resolution. This is in contrast and superior to the typical NGP methods.

In \cite{2014MNRAS.441.2923C} he effects of $\Rs$ on large-scale classifications was examined in detail. In this work, we mainly examine the influence of the grid resolution ($\rm N_{\rm grid}^{3}$). In forthcoming studies, we will further consider the effects of smoothing scale on large-scale classification algorithms, and the algorithm will be used for a variety of purpose, including the correlation between halo/galaxy properties and cosmic web, and the formation and evolution of haloes/galaxies within the cosmic web. This algorithm will also be used as the cornerstone for the precise identification of the filamentary structure.

\section*{Acknowledgments}
The authors acknowledge support from the joint Sino-German DFG research Project ``The Cosmic Web and its impact on galaxy formation and alignment'' (DFG-LI 2015/5-1). The work is supported by the NSFC (No.11825303, 11861131006, 11333008), the 973 program (No. 2015CB857003, No. 2013CB834900), and the NSF of Jiangsu Province (No. BK20140050). Q.G. acknowledges the sponsorship from Shanghai Pujiang Program 19PJ1410700. NIL acknowledges financial support of the Project IDEXLYON at the University of Lyon under the Investments for the Future Program (ANR-16-IDEX-0005).

%\bibliography{bibliografia}

\end{document}